\documentclass{pasj00}
\SetRunningHead{K. Saitou et al.}{Suzaku X-Ray Study of XSS\,J12270--4859}

\Received{2009/4/9}
\Accepted{2009/4/28}

\begin{document}
\title{Suzaku X-Ray Study of an Anomalous Source XSS\,J12270--4859}
\author{
Kei~\textsc{Saitou},\altaffilmark{1,2}
Masahiro~\textsc{Tsujimoto},\altaffilmark{1}
Ken~\textsc{Ebisawa},\altaffilmark{1}
and
Manabu~\textsc{Ishida}\altaffilmark{1}
}
\altaffiltext{1}{Japan Aerospace Exploration Agency, Institute of Space and
Astronautical Science\\ 3-1-1 Yoshino-dai, Sagamihara, Kanagawa 229-8510}
\altaffiltext{2}{Department of Astronomy, Graduate School of Science, 
The University of Tokyo, 7-3-1 Hongo, Bunkyo-ku, Tokyo 113-0033}
\email{ksaitou@astro.isas.jaxa.jp}
\maketitle

\begin{abstract}
We report the results of the Suzaku X-ray observation of XSS\,J12270--4859, one of 
the hard X-ray sources in the INTEGRAL catalogue. 
The object has been classified as an intermediate polar (IP) by optical spectra 
and a putative X-ray period of $\sim$860~s. 
With a 30~ks exposure of Suzaku, we obtained a well-exposed spectrum in the 
0.2--70~keV band. 
We conclude against the previous IP classification based on the lack of Fe 
K$\alpha$ emission features in the spectrum and the failure to confirm the 
previously reported X-ray period. 
Instead, the X-ray light curve is filled with exotic phenomena, including repetitive 
flares lasting $\sim$100~s, occasional dips with no apparent periodicities, spectral 
hardening after some flares, and bimodal changes pivoting between quiet and active 
phases. 
The rapid flux changes, the dips, and the power-law spectrum point toward the 
interpretation that this is a low-mass X-ray binary. 
Some temporal characteristics are similar to those in the Rapid Burster and 
GRO\,J1744--28, making XSS\,J12270--4859 a very rare object. 
\end{abstract}

%%%%%%%%%%%%%%%%%%%%%%%%%%%%%%%%%%%%%%%%%%%%%%%%%%%%%%%%%%%%
\section{Introduction}\label{s1}
%%%%%%%%%%%%%%%%%%%%%%%%%%%%%%%%%%%%%%%%%%%%%%%%%%%%%%%%%%%%
Intermediate polars (IPs) are a subclass of cataclysmic variables (CVs), which 
are interacting binaries of a moderately magnetized ($10^{5}-10^{7}$~G) white 
dwarf (WD) with a late-type dwarf or giant companion. 
Accreting material from the secondary forms a disk around the WD, which is 
truncated by the magnetic field in the primary's vicinity. 
Defining features of IPs are (1) optical and X-ray modulation due to the spin and 
orbital motions, (2) hard X-ray emission with Fe K$\alpha$ features arising 
from the accretion shock at the base of the accretion column, and (3) a complex 
profile of X-ray absorption \citep{patterson94,warner95,ezuka99,ramsay08}.

IPs occupy a very small fraction of CVs; about 50 confirmed IPs account only for 
$\sim$2\% of catalogued CVs \citep{downes01,barlow06}. 
Nevertheless, their particular brightness in the hard X-rays to soft $\gamma$-rays 
makes IPs to stand out when the sky is seen in these bands. 
The INTErnational Gamma-Ray Astrophysics Laboratory (INTEGRAL) serves very well 
for this purpose. 
A total of 21 CVs, mostly IPs, are listed in the third source catalogue by the 
INTEGRAL Soft Gamma-Ray Imager (ISGRI; \cite{bird07}). 
These include newly discovered samples, which was made possible by the ISGRI's 
positional accuracy of $\lesssim$1\arcmin\ and systematic optical spectroscopic 
follow-up studies \citep{masetti04}. 

XSS\,J12270--4859 (hereafter J12270) is one of such sources. 
Initially discovered in the Rossi X-ray Timing Explorer (RXTE) slew survey 
\citep{revnivtsev04}, the source was classified as an IP based on the H and He 
emission lines in the optical spectra \citep{masetti06}. 
This claim is also supported by a hint of $\sim$860~s X-ray periodicity in 
the RXTE data \citep{butters08}. 

Accumulating evidence, however, poses doubt on the IP classification. 
The Fe K$\alpha$ features remain undetected and the complex absorption profile 
is absent in the X-rays \citep{butters08,landi09}. 
Optical monitoring failed to confirm the reported X-ray period, but instead 
found flickering and possible phase changes with different temporal behaviors 
\citep{pretorius09}. 
All these features disagree with the IP nature of this source. 

In this Letter, we examine the previous IP classification of J12270 based on 
the new X-ray data with Suzaku. 
We present a well-exposed spectrum and a 30~ks light curve to test the Fe K$\alpha$ 
emission, and the modulation and the rapid changes in flux. 
We conclude against the IP classification and propose that J12270 is rather a 
low-mass X-ray binary (LMXB) full of anomalous temporal behaviors.

%%%%%%%%%%%%%%%%%%%%%%%%%%%%%%%%%%%%%%%%%%%%%%%%%%%%%%%%%%%%
\section{Observation and Data Reduction}\label{s2}
%%%%%%%%%%%%%%%%%%%%%%%%%%%%%%%%%%%%%%%%%%%%%%%%%%%%%%%%%%%%

\begin{figure*}[th]
  \begin{center}
    \FigureFile(174mm,84mm){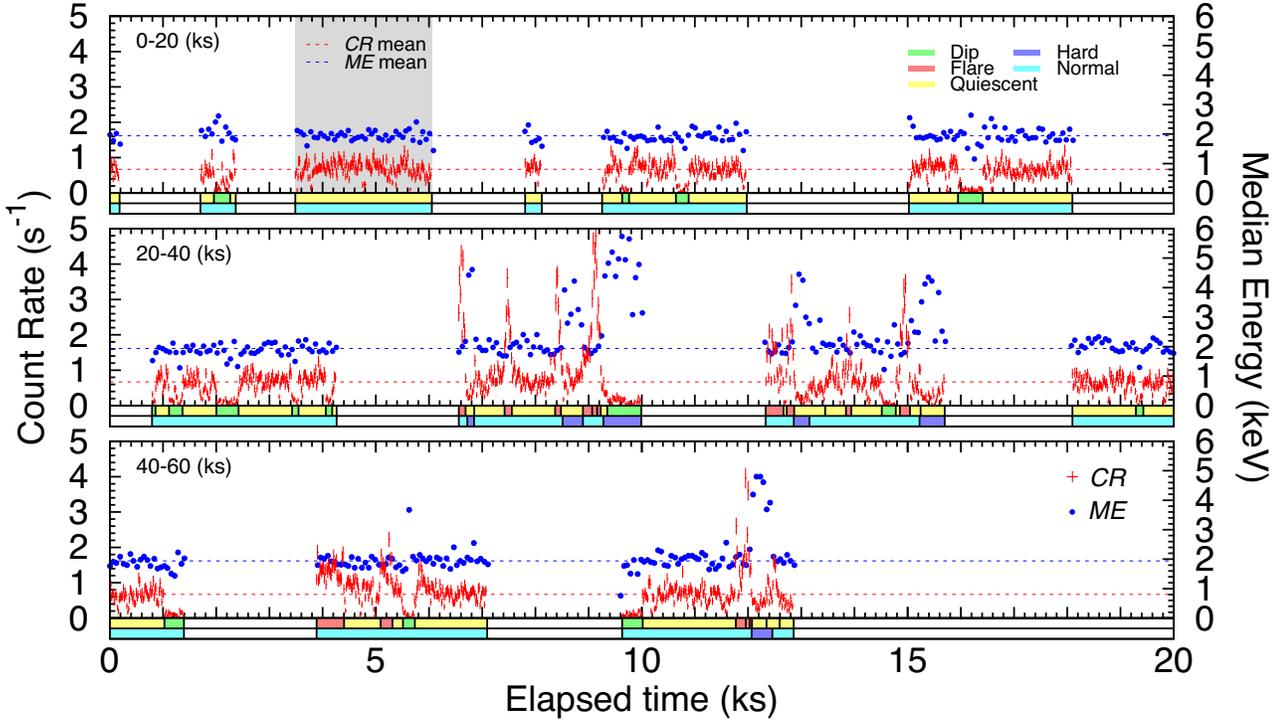}
  \end{center}
  \caption{XIS light curves of the background-subtracted \textit{CR} (red crosses) and 
           the \textit{ME} (blue bullets) in 0.2--12.0~keV. 
           The curves are folded by 20~ks. 
           The \textit{CR} and \textit{ME} curves are binned respectively with 16 and 
           64~s~bin$^{-1}$. 
           The origin of the abscissa is the observation start at 54686.97628~d in the 
           modified Julian date. 
           The 1$\sigma$ Poisson error is indicated for \textit{CR}. 
           The means of \textit{CR} and \textit{ME} determined in a quiescent interval 
           (gray shaded region) are represented by broken lines. 
           The two horizontal bars at the bottom of each panel represent segmented 
           blocks with different colors for dips, flares, and quiescence for \textit{CR} 
           and hard and normal for \textit{ME}.}
  \label{f1}
\end{figure*}

We observed J12270 with Suzaku (\cite{mitsuda07}) on 2008 August 8--9.  Suzaku has two
instruments: the X-ray Imaging Spectrometer (XIS; \cite{koyama07}) at 0.2--12~keV and
the Hard X-ray Detector (HXD; \cite{takahashi07}; \cite{kokubun07}) at 10--600~keV.

The XIS is equipped with four X-ray CCDs at the foci of four co-aligned X-ray 
telescopes (\cite{serlemitsos07}). 
One of them (XIS 1) is back-illuminated (BI) and the others (XIS 0, 2, and 3) are 
front-illuminated (FI) devices, which have an energy resolution of 150--190~eV (FWHM) 
at 5.9~keV. 
XIS 2 has been dysfunctional since 2006 November. 
Combined with the telescopes, the XIS has a total on-axis effective area of 
1030~cm$^2$ at 1.5~keV and a field of view (FoV) of 18\arcmin$\times$18\arcmin. 
The XIS was operated using the normal clocking with a frame time of 8~s. 

The HXD is a non-imaging detector comprised of several components covering different 
energy ranges. 
We focus on the PIN detector sensitive at 10--70~keV, which has an energy resolution 
of 3.0~keV (FWHM), a time resolution of 61~$\mu$s, and an effective area of 
$\sim$160~cm$^{2}$ at 20~keV. 
Passive fine collimators limit the FoV to $\sim$34\arcmin\ square in FWHM. 
The PIN achieves unprecedented sensitivity in this energy band due to the narrow FoV, 
the surrounding anti-coincidence detectors, and the low and stable background 
environment in a low-earth satellite orbit. 

The target at (RA, Dec) $=$ (\timeform{12h27m58s.9}, \timeform{-48D53'44''}) in the 
equinox J2000.0 was aimed at the center of the XIS field. 
The obtained data were processed with the standard pipeline version 2.2, leaving the 
net exposure time of 30~ks for the XIS and 35~ks for the HXD. 

For the XIS, we accumulated source and background signals for the temporal and spectral 
analyses. 
The source events were extracted from a 2\farcm95 radius circle around the object 
encompassing $\sim$90\% of photons, while the background events were from an annulus 
of 4\arcmin --7\arcmin \ in radii. 
The two XIS FI spectra with nearly identical responses were merged, while the BI 
spectrum was treated separately. 

For the PIN, we simulated background data for the spectral analysis. 
The PIN background is composed of instrumental non-X-ray background (NXB) and the 
cosmic X-ray background (CXB). 
The NXB spectrum was provided by the instrument team \citep{fukazawa09}, while 
the CXB spectrum was simulated by convolving the HEAO-1 model \citep{boldt87} with 
the detector responses. 
We found no contaminating source within the PIN FoV in the INTEGRAL catalogue 
\citep{bird07}.

%%%%%%%%%%%%%%%%%%%%%%%%%%%%%%%%%%%%%%%%%%%%%%%%%%%%%%%%%%%%
\section{Analysis}\label{s3}
%%%%%%%%%%%%%%%%%%%%%%%%%%%%%%%%%%%%%%%%%%%%%%%%%%%%%%%%%%%%

%%%%%%%%%%%%%%%%%%%%%%%%%%%%%%
\subsection{Light Curves}
%%%%%%%%%%%%%%%%%%%%%%%%%%%%%%

We constructed the binned light curves of J12270 (figure~\ref{f1}) after confirming 
that the background was non-variable. 
The background-subtracted count rate (\textit{CR}) and the median energy (\textit{ME}) 
were derived at 0.2--12.0~keV for each bin. 
\textit{ME} is defined as the median of energy of all photons in a bin, serving as 
a proxy for the spectral hardness suited for low photon statistics (\cite{hong04}). 

The object was occulted by the Earth for $\sim$1/3 of each 96~min orbit, causing 
discontinuities in the light curves. 
In the following, we call the continuous time spans in the light curves as ``intervals''. 
Variety of temporal features are noticeable at a glance. 
For example, in \textit{CR}, we can see sudden declines (we hereafter call these 
events ``dips'') and rapid amplification (``flares''). 
\textit{ME} is also variable, which shows significant hardening after some flares. 

For the quantitative definition of these features, we first divided the light curves 
into pieces of a constant \textit{CR} or \textit{ME} value (``segments'') using a 
Bayesian blocks (BB) method (\cite{scargle98}). 
The algorithm finds the change points of time-variable quantities, at which light 
curves are better explained by two different constant values before and after the 
point rather than a constant value. 
We next derived the base level of \textit{CR} and \textit{ME} using the third 
interval (gray shaded region in figure~\ref{f1}), during which no major variability 
is apparent. 
The mean and the standard deviation of \textit{CR} are 0.67 and 0.22~s$^{-1}$, 
respectively, while those of \textit{ME} are 1.94 and 0.17~keV, respectively. 
We finally defined dips as segments with \textit{CR} below the mean by $>$2$\sigma$, 
flares as segments with \textit{CR} above the mean by $>$2$\sigma$, and ``quiescence'' 
for the remaining \textit{CR} segments. 
Similarly, we defined ``hard'' segments with \textit{ME} above the mean by $>$2$\sigma$ 
and ``normal'' for the remaining \textit{ME} segments. 

As a result, we found 15 dips (a total of 4.2~ks) with no apparent periodicities and 11
flares (2.5~ks). All the six hard segments (2.5~ks) are preceded by flares.  The flares
are localized in some intervals, suggesting that the source has X-ray active phases
distinctive from quiet phases.

%%%%%%%%%%%%%%%%%%%%%%%%%%%%%%
\subsection{Timing}
\begin{figure}[h]
  \begin{center}
    \FigureFile(80mm,80mm){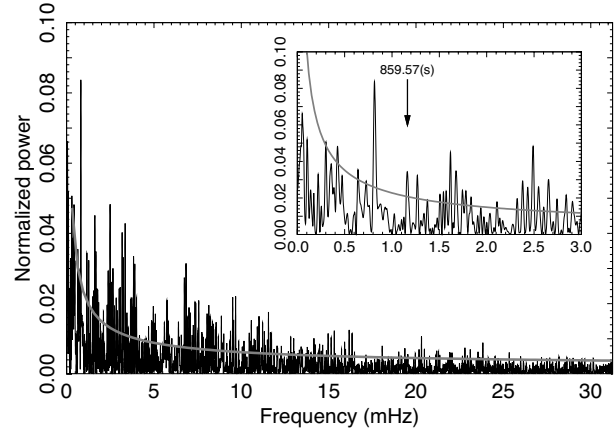}
  \end{center}
  \caption{Periodogram of 0.2--12.0~keV data in the quiescent segments. 
           The best-fit red noise profile is shown with the solid curve. 
           The white noise is estimated to be $\sim$0.002 in the power. 
           The inset shows a close-up view around the maximum and the previously 
           reported 860~s \citep{butters08} peaks.}
  \label{f2}
\end{figure}

%%%%%%%%%%%%%%%%%%%%%%%%%%%%%%
We employed a generalized Lomb-Scargle algorithm (GLS; \cite{zechmeister09}) 
for the timing analysis. 
GLS can process unevenly sampled data and takes the \textit{CR} errors fully 
into account, providing more accurate estimates of the frequency and power than 
conventional methods. 
We constructed periodograms up to the Nyquist frequency using selected segments 
and tested the maximum peak against the local noise. 
The red noise is dominant, which we derived by phenomenologically fitting with 
an exponential function \citep{vaughan05}. 
For the flare or dip segments, no significant peak above 3$\sigma$ was found, 
confirming their lack of periodicity. 
For all or the quiescent segments, similar negative results were obtained. 
Figure~\ref{f2} shows the result for the quiescent segments, in which we found 
the maximum power at $0.81$~mHz, not at the reported RXTE peak. 
The maximum is slightly above the 2$\sigma$ level, which we regard insignificant. 

\begin{figure}[h]
  \begin{center}
    \FigureFile(80mm,80mm){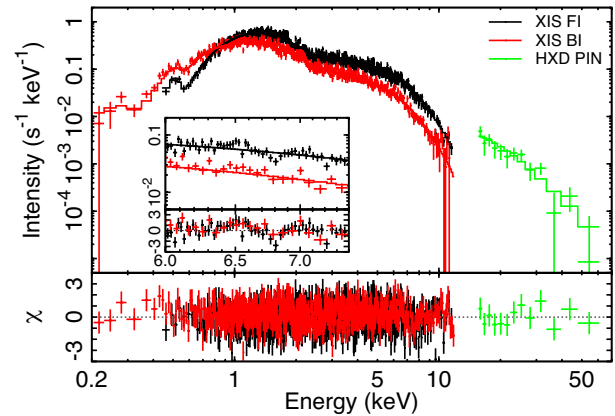}
  \end{center}
  \caption{Background-subtracted time-averaged XIS and PIN spectra. 
           The data and the best-fit power-law model are in the upper panel, while the 
           residuals are in the lower panel. 
           The enlarged view in 6.0--7.4~keV is in the inset.}
  \label{f3}
\end{figure}

%%%%%%%%%%%%%%%%%%%%%%%%%%%%%%
\subsection{Spectra}
%%%%%%%%%%%%%%%%%%%%%%%%%%%%%%
We constructed the background-subtracted spectra in the 0.2--70~keV band using the 
XIS and the PIN. 
We produced time-averaged spectrum (figure~\ref{f3}), which is overall featureless 
including the Fe K complex band (the inset). 
We also generated spectra for the stacked dip, flare, and hard segments. 
These spectra are quite similar to the time-averaged one, except for the decreased 
flux in the soft band for the dip and hard segment spectra. 

We fitted the XIS and PIN data simultaneously for the time-averaged and three sets 
of stacked spectra using a power-law or a thermal bremsstrahlung continuum convolved 
with an interstellar extinction (\texttt{tbabs}; \cite{wilms00}). 
All best-fits were acceptable (tables~\ref{t1} and \ref{t2}) except for unconstrained 
bremsstrahlung model for the dip and the hard spectra. 
Upon the best-fit continuum models, we added a Gaussian model for the tantalizing 
emission features at 6.5 and 7.0~keV and found that these features are  insignificant 
at 90\% confidence in the F test. 

\begin{table}[h]
  \begin{center}
    \caption{Best-fit parameters for the power-low model\footnotemark[$*$]}
    \label{t1}
  \begin{tabular}{lcccc}
    \hline
    State & $N_\mathrm{H}$ & $\Gamma$ & $F_\mathrm{X}$ & Re-$\chi^{2}$ (dof) \\ 
    \hline
    Average \dotfill 
      & $1.0^{+0.1}_{-0.1}$ 
      & $1.53^{+0.02}_{-0.02}$ 
      & $1.82^{+0.02}_{-0.02}$ 
      & 0.99 (1244) \\ 
    Dip \dotfill 
      & $<0.3$ 
      & $1.32^{+0.09}_{-0.07}$ 
      & $0.37^{+0.03}_{-0.02}$ 
      & 1.12 (\phantom{00}98) \\ 
    Flare \dotfill 
      & $1.2^{+0.2}_{-0.2}$ 
      & $1.62^{+0.04}_{-0.04}$ 
      & $4.54^{+0.11}_{-0.11}$ 
      & 0.90 (\phantom{0}603) \\ 
    Hard \dotfill 
      & $2.9^{+1.3}_{-1.3}$ 
      & $0.64^{+0.10}_{-0.10}$ 
      & $1.70^{+0.27}_{-0.27}$ 
      & 1.28 (\phantom{0}141) \\ 
   \hline
   \multicolumn{5}{@{}l@{}}{\hbox to 0pt{\parbox{90mm}{\footnotesize 
   \par\noindent 
   \footnotemark[$*$] The parameters are the absorption column density ($N_\mathrm{H}$) 
                      in the unit of 10$^{21}$~cm$^{-2}$, the photon index ($\Gamma$), 
                      and the 0.2--12.0~keV flux ($F_{\mathrm{X}}$) in 
                      10$^{-11}$~erg~s$^{-1}$~cm$^{-2}$. 
                      The ranges of parameter values indicate the 90\% statistical uncertainty. 
                      The goodness of the fit is indicated with the reduced $\chi^{2}$ value 
                      (Re-$\chi^{2}$) and the degree of	freedom (dof). 
    }\hss}}
  \end{tabular}
  \end{center}
\end{table}

\begin{table}[h]
  \begin{center}
    \caption{Best-fit parameters for the bremsstrahlung model\footnotemark[$*$]}
    \label{t2}
  \begin{tabular}{lcccc}
    \hline
    State & $N_\mathrm{H}$ & $k_\mathrm{B}T$ & $F_\mathrm{X}$ & Re-$\chi^{2}$ (dof) \\ 
    \hline
    Average \dotfill 
      & $0.6^{+0.1}_{-0.1}$ 
      & $16.4^{+1.0}_{-0.9}$ 
      & $1.77^{+0.11}_{-0.10}$ 
      & 1.00 (1244) \\ 
    Flare \dotfill 
      & $0.6^{+0.1}_{-0.1}$ 
      & $11.6^{+1.1}_{-1.0}$ 
      & $4.37^{+0.41}_{-0.38}$ 
      & 0.93 (\phantom{0}603) \\ 
    \hline
    \multicolumn{5}{@{}l@{}}{\hbox to 0pt{\parbox{90mm}{\footnotesize 
    \par\noindent 
    \footnotemark[$*$] The notations follow table~\ref{t1} except for the electron 
                       temperature ($k_\mathrm{B}T$) in keV.
    }\hss}}
  \end{tabular}
  \end{center}
\end{table}

%%%%%%%%%%%%%%%%%%%%%%%%%%%%%%%%%%%%%%%%%%%%%%%%%%%%%%%%%%%%
\section{Discussion}\label{s4}
%%%%%%%%%%%%%%%%%%%%%%%%%%%%%%%%%%%%%%%%%%%%%%%%%%%%%%%%%%%%
Our results show that J12270 satisfies none of the defining X-ray characteristics 
of IPs. 
First, Fe K$\alpha$ emission is ubiquitously seen not only in magnetic CVs 
\citep{ezuka99,demartino04,suleimanov05} but also in non-magnetic CVs such as dwarf 
novae \citep{pandel05,rana06}. 
However, we found no evidence of this emission for J12270. 
The [Fe/H] abundance is $<$0.14~solar (90\%), which is below the typical range 
(0.2--0.6) for magnetic CVs \citep{ezuka99}. 
Second, for the reported 860~s period in \citet{butters08}, we retrieved their RXTE 
data and found that the light curve exhibits flares and dips similarly to Suzaku. 
We confirmed the reported period in the periodogram using all data, but it was 
rendered insignificant after removing flares and dips. 
Third, the occasional flares and the pivots between the quiet and active phases 
in J12270 is uncommon for IPs except for AE Aqr \citep{choi99}. 
However, the time scales of these behavior in AE Aqr are much longer than J12270. 
With all the counter-evidence, we conclude that J12270 is not an IP. 

The short and aperiodic changes in the X-ray and optical flux suggest that J12270 
constitutes an accreting system. 
The lack of early-type signatures in the unobscured optical spectrum \citep{masetti06} 
indicates that the secondary is a late-type star. 
The flux changes in $\sim$100~s and the power-law spectrum point toward a LMXB nature 
of this source. 
In fact, LMXBs commonly exhibit short bursts and dips. Bursts can be caused by 
thermonuclear flashes on the neutron star surface (type I) or the sudden increase of 
the accretion rates due to disk instability (type II). 
The flares in J12270 are quite similar to those in the Rapid Burster \citep{lewin76}, 
which is the prototypical type II burster, or GRO\,J1744--28 \citep{kouveliotou96}. 
The flares in these sources commonly show the repetition of short flares, the flux 
amplification by a factor of $\sim$5 from the quiet level, and the flux decrease 
immediately after some flares. 

The distance to J12270 was estimated as $\sim$220~pc \citep{masetti06} assuming that 
the \textit{V}-band flux is dominated by the secondary. 
However, the variability with a large amplitude of $>$1~mag in this band 
\citep{pretorius09} indicates that the emission from the accreting material accounts 
for a substantial fraction. 
Indeed, a ten-fold increase in the distance estimate can be easily reconciled with 
observations. 
If the source is located at 2.2~kpc, the X-ray luminosity is 
$\sim$1$\times$10$^{34}$~erg~s$^{-1}$, and the height above the Galactic plane 
is 0.5~kpc, which exceeds the scale height of the Galactic interstellar 
H\emissiontype{I} gas \citep{spitzer04}. 
The integrated H\emissiontype{I} column of 1$\times$10$^{21}$~cm$^{-2}$ in this 
direction \citep{kalberla05} is consistent with our spectral fitting result 
(table \ref{t1}). 
Both the X-ray luminosity and the scale height estimated for 2.2~kpc are reasonable 
for a LMXB \citep{christian97}.

%%%%%%%%%%%%%%%%%%%%%%%%%%%%%%%%%%%%%%%%%%%%%%%%%%%%%%%%%%%%
% Acknowledgment
%%%%%%%%%%%%%%%%%%%%%%%%%%%%%%%%%%%%%%%%%%%%%%%%%%%%%%%%%%%%
\bigskip

The authors thank J.~D.~Scargle and M.~Zechmeister for providing the BB and GLS 
scripts, respectively, and K.~Mukai, A.~Imada and T.~Dotani for helpful comments.

%%%%%%%%%%%%%%%%%%%%%%%%%%%%%%%%%%%%%%%%%%%%%%%%%%%%%%%%%%%%
% Bibliography
%%%%%%%%%%%%%%%%%%%%%%%%%%%%%%%%%%%%%%%%%%%%%%%%%%%%%%%%%%%%

\end{document}